\documentclass[aps,prl,onecolumn]{revtex4-2}
\usepackage[utf8]{inputenc}

\usepackage{bm}
\usepackage{amssymb}
\usepackage{mathtools}
\usepackage{xcolor}
\usepackage{enumitem}
\usepackage{natbib}
\usepackage{hyperref}
\usepackage{lipsum}
\usepackage[ruled,vlined]{algorithm2e}

\usepackage{amsmath}

\begin{document}
\title{Assessing the impact of Byzantine attacks on coupled phase oscillators}
\author{Melvyn Tyloo}
\address{Theoretical Division and Center for Nonlinear Studies (CNLS), Los Alamos National Laboratory, Los Alamos, NM 87545, USA}
\begin{abstract}
For many coupled dynamical systems, the interaction is the outcome of the measurement that each unit has of the others as e.g. in modern inverter-based power grids, autonomous vehicular platoons or swarms of drones, or it is the results of physical flows. Synchronization among all the components of these systems is of primal importance to avoid failures. The overall operational state of these systems therefore crucially depends on the correct and reliable functioning of the individual elements as well as the information they transmit through the network. Here, we investigate the effect of Byzantine attacks where one unit does not behave as expected, but is controlled by an external attacker. For such attacks, we assess the impact on the global collective behavior of nonlinearly coupled phase oscillators. We relate the synchronization error induced by the input signal to the properties of the attacked node. This allows to anticipate the potential of an attacker and identify which network components to secure.
\end{abstract}
\maketitle

\section{Introduction}

Networked systems where the interaction among the individual units is mediated through the perception each units have of its neighbors or via the exchange of a physical quantity find numerous realizations both in natural and engineered systems~\cite{Pik03,strogatz2014nonlineardynamics}.  For example, traditional power plants having a rotating mass typically synchronize thanks to the flows of power they are exchanging, while renewable energy sources such as solar panels or wind turbines usually read the voltage frequency on the grid and inject their power at this particular frequency~\cite{pattabiraman2018comparison}. Another example are autonomous cars and drones being part of a larger vehicular platoon or swarm, where each component measures the distance between itself and its neighbors~\cite{8778746}. From this interaction, collective macroscopic phenomena emerge such as synchronization or consensus. Such coherent operational states are made possible thanks to the interplay between both the units' internal parameters and the coupling within them. Therefore, alterations or attacks of either the coupling within the interacting units or the units themselves might dramatically impact the whole state of the system and, thus, its good and desired functioning.
Robustness of network-coupled dynamical system can be investigated from various points of view. One may consider the effect of input perturbation signals or noise on the overall stability of the system~\cite{Dev12,bamieh2013price,Pag17,doi:10.1063/5.0129123,Hin18}, or assess the vulnerabilities to line and node failure or removal~\cite{DELABAYS2022270}, and how such attacks may lead to cascades where, eventually, the operational state of the system is disrupted~\cite{crucitti2004model,PhysRevE.95.012315}. Another way to impact the stability of the system is by spoofing some components. Indeed, the collective state characterizing the stable state of coupled dynamical system essentially depends on the relative values of the degrees of freedom of each and every interacting elements~\cite{strogatz2014nonlineardynamics,Str04}. Therefore, if one or a subset of these degrees of freedom can be controlled or spoofed, then one may globally impact the collective behavior of the system and, potentially, disrupt it~\cite{8320800}. In this manuscript, we investigate the latter scenario where an attacker has the ability to control a single degree of freedom, i.e. a single unit, of a system of networked oscillators, which is often referred to as a \textit{Byzantine attack}~\cite{lamport1982byzantine}. Such situation could take place, for example, in the formation of vehicular platoons or swarming robots where some attacker controls some sensors or the shared information. This problem has been previously investigated for pulse-coupled oscillators where Ref.~\cite{8320800} showed that synchronization is still achievable despite Byzantine type of attacks. The resilience to such malicious attacks of discrete consensus algorithm including mobile agent has been explored in Ref.~\cite{8013830}. Detection and mitigation of false data injection in controlled networked systems in general, and microgrids has recently attracted an increasing interest~\cite{10114521855052185515,8260848,8894512}. Less has been done on the effect of Byzantine attack on the collective behavior of phase-coupled oscillators, which are directly related to integrator dynamics describing e.g. vehicular platoon formation~\cite{li2015overview}.

Here we investigate the global impact on synchronization induced by an attacker, who has the full control of a single oscillator. Based on response theory, we develop a framework to investigate the reaction of the system to such attacks and use it to calculate a performance metric that assess how synchronization is affected by an external perturbation. We derive a general expression for arbitrary input signals the attacker would choose and consider then more precisely the two cases of random and periodic signals for which the properties of the attacked node can be directly connected to the disruptive potential of the attack.

The manuscript is organized as follows. Section~\ref{sec1} gives the mathematical notations. Section~\ref{Sec1} introduces the networked oscillator model and defines Byzantine attacks we consider. Section~\ref{sec2} gives the response theory of the system to external attacks and defines the performance metric to assess the potential of the attack. Section~\ref{sec_rand} applies the theory to the specific cases of random and periodic input signals. Section~\ref{Num} illustrates numerically the theory developed in the previous sections.
Conclusions are given in Section~\ref{conc}.

\begin{figure}
    \centering
    \includegraphics[scale=0.7]{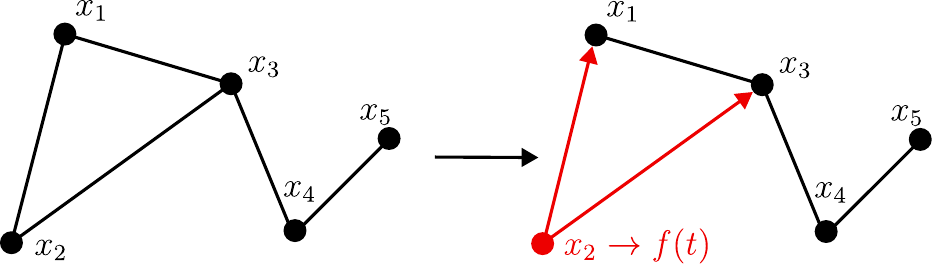}
    \caption{Left: original network before the attack; right: network following the attack where the degree of freedom of oscillator 2 is replaced by an input signal $f(t)$\,. The resulting system is an oscillatory network made of nodes 1, 3, 4, 5 where oscillators 1 and 3 are influenced by the input signal from the attacker, i.e. $1,3\in \mathcal{N}(2)$\,.}
    \label{fig1}
\end{figure}

\section{Mathematical notations}\label{sec1}
We consider a network of $N$ nodes where the $k$-th node is controlled by an attacker. The set of nodes connected to node $k$ is denoted by $\mathcal{N}(k)$\,. 
In the following, we write column vectors ${\bf v}\in \mathbb{R}^N$ as bold lower case letters. Matrices ${\bf M}\in \mathbb{R}^{(N\times N)}$ are denoted with bold upper case letters. The Kroenecker symbol equivalent to the $j$-th component of the unit vector is written as $\hat{e}^i_j=\delta_{ij}$\,. The Dirac delta function is denoted $\delta(t)$\,.

\section{Byzantine attacks on Networked Oscillators}\label{Sec1}
\subsection{Kuramoto oscillators}
As a paradigmatic model for investigating synchronization phenomena, we consider $N$ coupled Kuramoto oscillators whose time-evolution is governed by the following set of coupled differential equations~\cite{Kur75},
\begin{eqnarray}\label{eq1} 
\dot{\theta}_i = \omega_i - \sum_j b_{ij}\sin(\theta_i-\theta_j) \,,i=1,...N\,,
\end{eqnarray}
where the degrees of freedom are defined over a compact set i.e. $\theta_i\in[0,2\pi)$\,, and $\omega_i$ is the natural frequency of the $i$-th oscillator. The coupling between the oscillators is given by the elements of the adjacency matrix $b_{ij}$\,. If the natural frequencies are not too broadly distributed and the network is connected, with the coupling that is strong enough, the oscillators eventually reach a synchronized state (also called \textit{phase-locked state}) such that $\dot{\theta}_i(t\rightarrow\infty)=\Omega=\sum_i N^{-1}\omega_i$ $\forall i$\,. Without loss of generality, one can move to the rotating frame and thus have $\Omega =0$\,. Depending on the coupling topology and the natural frequencies, many stable synchronous states may exist, each of them with their own basin of attraction. Perturbations applied to the system can induce transitions between the different synchronous states or even simply disrupt synchrony~\cite{Dev12,Tyl18c,Hin18}. 
\subsection{Byzantine attack}
The type of attack we consider in this manuscript is inspired by the Byzantine Generals problem~\cite{lamport1982byzantine}, where an individual being part of a larger networked game, decides to pass some corrupted information to their neighbors, instead of behaving reliably as expected. In our framework of coupled oscillators, we model this as an attacker having full control of an oscillator $k$\,, which is formulated as,
\begin{eqnarray}\label{eq2}
    \tilde\theta_k(t) = \theta_k(t) + f(t) \,, t>0\,
\end{eqnarray}
where $f(t)$ is the input signal that is chosen by the attacker, which might be e.g. constrained by some budget, and $ \tilde\theta_k$ is the controlled degree of freedom of node $k$\,. The controlled input signal Eq.~(\ref{eq2}) directly affect the neighbors of node $k$ whose time-evolution is then given by,
\begin{eqnarray}\label{eq3} 
\dot{\theta}_{i} = \omega_i - \sum_{j\neq k} b_{ij}\sin(\theta_i-\theta_j) - b_{ik}\sin[\theta_i- \tilde\theta_k]  \,,
\end{eqnarray}
where $b_{ik}$ is non-vanishing if $i\in \mathcal{N}(k)$ with $\mathcal{N}(k)$ the set of the neighbors of node $k$\,. An important question is then, how robust is the synchronous state to such attacks. One must notice that Eq.~(\ref{eq3}) effectively changes the coupling network topology by removing node $k$\, as the latter is not influenced by its neighbors anymore. The situation is illustrated on Fig.~\ref{fig1}. In the following, we investigate this question using the synchronization error and the linear response of the system.

\section{Assessment of the system's response}\label{sec2}
\subsection{Linear response}
Before the attack (i.e $t<0$), we assume that the system sits at a synchronous state $\{\theta_i^{(0)}\}$ that satisfies the algebraic equations,
\begin{eqnarray}\label{eq40} 
0 = \omega_i - \sum_j b_{ij}\sin(\theta_i^{(0)}-\theta_j^{(0)}) \,,i=1,...N\,.
\end{eqnarray}
It is clear that many stable fixed point might exist or even none at all, because of the nonlinear coupling function. Here we assume that at least one stable synchronous solution to Eq.~(\ref{eq40}) exists before the attack. Further assuming that the perturbations applied to the system are small enough, the behavior of the oscillators following the attack is well approximated by their linear response. Defining the deviation from the synchronous state $\phi_i(t) = \theta_i(t)-\theta_i^{(0)}$ for $i\neq k$ and $\phi_k(t) = f(t)$ and taking into account the attack Eq.~(\ref{eq2}) one has the dynamics,
\begin{align}\label{eq4} 
\begin{split}
\dot{\phi}_i =
\begin{cases}
- \sum_j \tilde{\mathbb{L}}_{ij}\,\phi_j  \quad \text{for }i\neq k \,,i\notin \mathcal{N}(k)\,,\\
-\sum_{j\neq k} \tilde{\mathbb{L}}_{ij}\,\phi_j - b_{ik}\cos(\theta_i^{(0)}-\theta_k^{(0)})[\phi_i-f(t)]\, \\ \quad \text{for }i\neq k \,,i\in \mathcal{N}(k)\,\\
\dot{f}(t) \quad \text{for } i=k\,,
\end{cases}
\end{split}
\end{align}
where we defined the weighted Laplacian matrix $\tilde{\mathbb{L}}\in \mathbb{R}^{(N-1)\times(N-1)}$ of the network where node $k$ has been removed (see Fig.~\ref{fig1}),
\begin{eqnarray}\label{eq6}
   \tilde{\mathbb{L}}_{ij} = 
   \begin{cases}
           -b_{ij}\cos(\theta_i^{(0)}-\theta_j^{(0)}) \,, i\neq j\,,\\
            \sum_{k}b_{ik}\cos(\theta_i^{(0)}-\theta_k^{(0)}) \,, i=j\,.
   \end{cases}
\end{eqnarray}
This matrix depends on both the coupling network and the synchronous state of the system. Note that, in the rest of this manuscript we denote the effective weights on the edges of the new network $\tilde{b}_{ij}=b_{ij}\cos(\theta_i^{(0)}-\theta_j^{(0)})$\,.
Rather interestingly, the dynamics given by Eq.~(\ref{eq4}) corresponds to the Taylor model that describes the interaction between networked agents where some of them are stubborn i.e. have an additional bias~\cite{doi:10.1177/001872676802100202, Bau20}. 
Restricting Eq.~(\ref{eq4}) to the uncontrolled oscillators by defining the new variable $\bf \varphi$ which is simply $\bf \phi$ where the $k$-th component has been removed, one can write the dynamics in a matrix form as,
\begin{eqnarray}\label{eq7}
    \dot{{\bf \varphi}} = - (\tilde{\mathbb{L}} + {\bf K})\,{\bf \varphi} + {\bf{g}}(t)  \,,
\end{eqnarray}
where we defined the diagonal matrix with non-zero elements given by $K_{ii} = \tilde{b}_{ik}$ for $i\in \mathcal{N}(k)$\,, and the vector $\bf g$ with non-zero components ${g}_i(t) = \tilde{b}_{ik}\,f(t)$ for $i\in\mathcal{N}(k)$\,. Equation~\ref{eq7} is a linear system which can be solved by expanding the deviations over the eigenbasis of the symmetric matrix $(\tilde{\mathbb{L}} + {\bf K})$\,, i.e. $\varphi_i(t)=\sum_{\alpha}c_\alpha(t)\,u_{\alpha,i}$\, where $c_\alpha(t)=\sum_j \varphi_j(t)u_{\alpha,j}$\,. The eigenvalues of $(\tilde{\mathbb{L}} + {\bf K})$ satisfy $\lambda_k>0$ and set the time-scales of the response dynamics. The corresponding eigenvectors are denoted ${\bf u}_{\alpha}$\,. The time-evolution of the system is then given by,
\begin{eqnarray}\label{eq8}
    \varphi_i(t) = \sum_\alpha e^{-\lambda_\alpha t}\int_0^t e^{\lambda_\alpha t' } \sum_j g_j(t')u_{\alpha,j}\, {\rm d}t'\,u_{\alpha,i}\,.
\end{eqnarray}
The latter equation gives the response of the system following an attack at node $k$ where its degree of freedom has been replaced by $f(t)$\,. In the following section, we introduce a metric to assess the impact of the attack on the overall synchronization of the system, which can be calculated using Eq.~(\ref{eq8}).

\subsection{Synchronization error}
In order to quantify the global impact of the attack on the synchronization of the system, we consider the so-called \textit{synchronization error} which is a common performance metric for networked systems subjected to perturbations~\cite{bamieh2013price,Pag17,Tyl18a,Bau20}. We define the synchronization error at time $t$ as,
\begin{eqnarray}
    \mathcal{P}(t) = \sum_{i<j}\tilde{b}_{ij}[\varphi_i(t) - \varphi_j(t)]^2\,.
\end{eqnarray}
It measures the mismatch in the angle deviations between neighboring oscillators, due to the attack. Large values of $\mathcal{P}$ mean that the deviations between connected oscillators were asynchronous while small values correspond to a synchronous response of the system where most of the deviations followed the same direction. Moreover, it can be conveniently written as,
\begin{eqnarray}\label{eq9}
    \mathcal{P}(t) &=& \sum_{i,j}\varphi_i(t)\,\tilde{\mathbb{L}}_{ij}\,\varphi_{j}(t)= \sum_{\alpha}\lambda_\alpha\,c_\alpha^2(t) - \sum_{j\in\mathcal{N}(k)}\tilde{b}_{jk}\varphi_j^2(t)\,,
\end{eqnarray}
where in the second equality we used the expansion of the deviations over the eigenbasis and the orthogonality between the eigenvectors ${\bf u}_\alpha$'s. The synchronization error can be obtained for any input signal from the attacker by plugging $f(t)$ into Eq.~(\ref{eq8}) and then substituting it in Eq.~(\ref{eq9}). One can therefore investigate any desired signal restricted or not to some budget which models the limited capacities of the attacker. Here, to get an intuitive picture of the potential of the attack, we consider a random and a periodic input signal.
\begin{figure}
    \centering
    \includegraphics[scale=0.7]{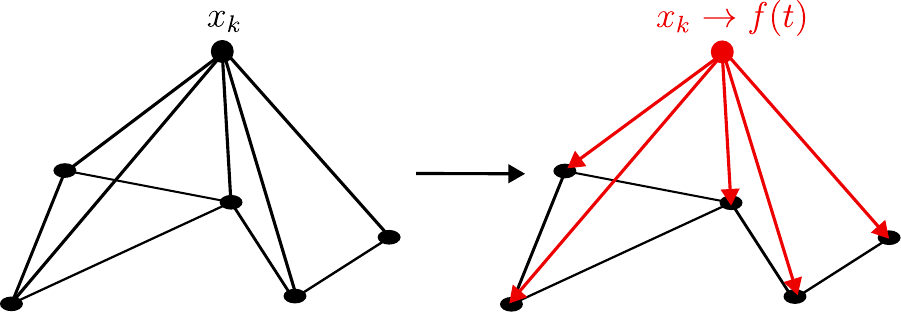}
    \caption{Left: original network before the attack; right: network following the attack where the degree of freedom of oscillator $k$, which is connected to all other nodes, is replaced by an input signal $f(t)$\,.}
    \label{fig2}
\end{figure}

\section{Input signals}\label{sec_rand}
\subsection{White noise}
The simplest input the attacker could choose is that of a random signal around the initial value of the degree of freedom at node $k$\,, i.e. $\langle f(t) \rangle = 0$\,, where $\langle . \rangle$ denotes the statistical average which also corresponds to the time average i.e. $\langle f(t) \rangle = \lim_{T\rightarrow\infty}T^{-1}\int_0^T f(t'){\rm d}t'$\,. If the latter is not correlated in time one has
\begin{eqnarray}\label{eq10}
    \langle f(t) f(t') \rangle = \tau_0 \, \delta({t-t'})\,,
\end{eqnarray}
where $\tau_0$ is a time constant introduced to keep the input signal $f(t)$ without units and which also gives the standard deviation of the signal. Using this two-point correlator, one can then calculate the average of the synchronization error. 
Plugging Eq.~(\ref{eq10}) into Eq.~(\ref{eq9}) yields,
\begin{align}
\begin{split}
    \langle \mathcal{P} \rangle = \lim_{t\rightarrow\infty}\left[\sum_\alpha \lambda_\alpha e^{-2\lambda_\alpha t}\right.&\int \hspace{-0.2cm}\int_0^t e^{\lambda_\alpha (t_1+t_2)}  \hspace{-0.2cm}\sum_{i,j\in\mathcal{N}(k)}\tilde{b}_{ik}\tilde{b}_{jk}u_{\alpha,i}u_{\alpha,j}  \langle f(t_1)f(t_2) \rangle{\rm d}t_1{\rm d}t_2\\
      - \sum_{\alpha,\beta}  e^{-(\lambda_\alpha + \lambda_\beta) t}\int \hspace{-0.2cm}\int_0^t & e^{\lambda_\alpha t_1}e^{\lambda_\beta t_2} \hspace{-0.2cm} \sum_{l,m\in\mathcal{N}(k)}\tilde{b}_{lk}\tilde{b}_{mk}u_{\alpha,l}u_{\beta,m} \langle f(t_1)f(t_2) \rangle {\rm d}t_1{\rm d}t_2 \left.\sum_{j\in\mathcal{N}(k)}\tilde{b}_{jk}u_{\alpha,j}u_{\beta,j}  \right]\,,
     \end{split}
\end{align}
which, after some algebra gives
\begin{align}\label{eq11}
\begin{split}
     \langle \mathcal{P} \rangle = \frac{\tau_0}{2}\sum_{j\in\mathcal{N}(k)} \tilde{b}^2_{jk} - \tau_0\sum_{\alpha,\beta}\sum_{i,j,l\in\mathcal{N}(k)}\tilde{b}_{ik}\tilde{b}_{jk}\tilde{b}_{lk}\frac{u_{\alpha,i}u_{\beta,j}u_{\alpha,l}u_{\beta,l}}{\lambda_\alpha+\lambda_\beta}\,,
     \end{split}
\end{align}
\begin{figure}
    \centering
    \includegraphics[scale=0.7]{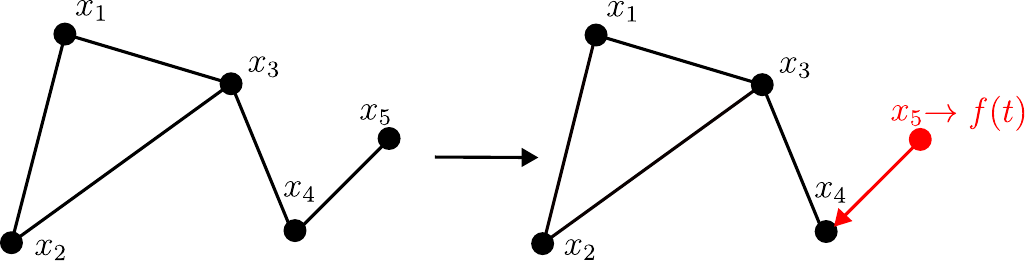}
    \caption{Left: original network before the attack; right: network following the attack where the degree of freedom of oscillator $k$, which is connected to a single other node, is replaced by an input signal $f(t)$\,.}
    \label{fig20}
\end{figure}
where we recall that ${\bf u}_\alpha$ and $\lambda_\alpha$ are respectively the eigenvectors and eigenvalues of the matrix $(\tilde{\mathbb{L}} + {\bf K})$\,.
While the first term in the synchronization error Eq.~(\ref{eq11}) is simply proportional to the sum of the squared edge weights of the attacked node $k$ in the initial network, the second term is more intricate with the eigenmodes and the weights between the attacked node and its neighbors. The latter might becomes large when the nodes connected to $k$ are sitting on the slowest eigenmodes for which $\lambda_\alpha$ is small. The synchronization error for such random input signal is therefore a trade-off between a local property of the system given by the first term in Eq.~(\ref{eq11}) and a more global one given by the second term. It is important to notice that the synchronization error for this Byzantine attack Eq.~(\ref{eq11}) is different from what one would get by having an additive white-noise perturbation at oscillator $k$~\cite{Tyl19}. In the next section, we illustrate numerically the theoretical predictions.

At this stage, it is instructive to briefly discuss the case where the attacked node $k$ is connected to all the nodes in the network. This would happen in a system where all units are connected to an oscillator possibly modelling a centralized control unit. This is illustrated in Fig.~\ref{fig2}. Then, $\mathcal{N}(k)$ includes all the network nodes. Further assuming that the weights between node $k$ and the rest of the network are the same i.e. $\tilde{b}_{ik}=b$ $\forall i$\,, one has for the synchronization error $\langle \mathcal{P} \rangle=0$\,.
Quite intuitively, attacking the oscillator that is connected to all the other ones in the network results in injecting the same signal to every node and thus, all nodes remain synchronized. Therefore, attacking such node is not effective if its coupling to all the other nodes is homogeneous. If it is heterogeneous, one expects a non-vanishing synchronization error.

The opposite of the latter situation would be an attack on a node that has a single connection to the rest of the network, as shown in Fig.~\ref{fig20}. In such a scenario, Eq.~(\ref{eq11}) becomes,
\begin{align}
\begin{split}
     \langle \mathcal{P} \rangle = \frac{\tau_0}{2} b^2 - \tau_0\,b^3\,\sum_{\alpha,\beta}\frac{u_{\alpha,i}^2u_{\beta,i}^2}{\lambda_\alpha+\lambda_\beta}\,,\,\,\, \text{with}\,\,\, i\in \mathcal{N}(k)\,.
     \end{split}
\end{align}
In this case, the effect of the Byzantine attack is limited as the synchronization error is small than $\frac{\tau_0}{2} b^2$\,

\subsection{Periodic signal}
Instead of using a stochastic input signal, the attacker might use a deterministic function for $f(t)$\,. Here, we calculate the synchronization error when the attacker chooses a periodic signal given by,
\begin{eqnarray}\label{forc}
    f(t) = \gamma\, \cos(\omega\,t)\,,
\end{eqnarray}
with $\omega$ the angular frequency and $\gamma$ the amplitude of the input. To shorten the notation, we set $\gamma=1$ in the following. In the long time limit, one has,
\begin{align}\label{per}
\begin{split}
    \mathcal{P}(t) &= \sum_\alpha\sum_{i,j\in\mathcal{N}(k)} \tilde{b}_{ik}\tilde{b}_{jk} u_{\alpha,i}u_{\alpha,j} \frac{\lambda_\alpha [\lambda_\alpha^2+\omega^2 + (\lambda_\alpha^2-\omega^2)\cos(2\omega\,t) + 2\lambda_\alpha\omega\sin(2\omega\,t) ]}{2(\lambda_\alpha^2 + \omega^2)^2}\\
    &\hspace{-1cm}- \sum_{\alpha,\beta}\sum_{i,j,l\in\mathcal{N}(k)} \tilde{b}_{ik}\tilde{b}_{jk}\tilde{b}_{lk}  u_{\alpha,i}u_{\beta,j}u_{\alpha,l}u_{\beta,l}\frac{\lambda_\alpha\lambda_\beta + \omega^2 + [\lambda_\alpha\lambda_\beta-\omega^2]\cos(2\omega\,t) + \omega[\lambda_\alpha + \lambda_\beta]\sin(2\omega\,t)  }{2(\lambda_\alpha^2 + \omega^2)(\lambda_\beta^2 + \omega^2)}\,.
    \end{split}
\end{align}
To gain more intuition about the latter expression, we consider the two limiting case of $\lambda_N/\omega \ll 1$ and $\lambda_2/\omega\gg 1$\,. In the first case of a high-frequency attack, one has,
\begin{align}\label{hf}
\begin{split}
    \mathcal{P}(t) &= \left[\sum_{i,j\in\mathcal{N}(k)} \tilde{b}_{ik}\tilde{b}_{jk} \tilde{\mathbb{L}}_{ij} \right]\frac{[1 -\cos(2\omega\,t) ]}{2\,\omega^2}\,, \,\, \lambda_N/\omega \ll 1 \,.
    \end{split}
\end{align}
In this limit, the synchronization error is deduced from the local network structure around $k$. This is expected as high-frequency disturbances do not spread across the system. A larger response is predicted when the neighbors of the attacked node $k$ are not directly connected. In the other limit of a low-frequency attack, one has,
\begin{align}\label{lf}
\begin{split}
    \mathcal{P}(t) &= \sum_{i,j\in\mathcal{N}(k)} \tilde{b}_{ik}\tilde{b}_{jk} [\tilde{\mathbb{L}} + {\bf K}]^{-1}_{ij}  \frac{[1 + \cos(2\omega\,t) ]}{2}\\
    &- \sum_{i,j,l\in\mathcal{N}(k)} \tilde{b}_{ik}\tilde{b}_{jk}\tilde{b}_{lk}  [\tilde{\mathbb{L}} + {\bf K}]^{-1}_{il}[\tilde{\mathbb{L}} + {\bf K}]^{-1}_{jl}\frac{[1  + \cos(2\omega\,t) ]}{2}\,, \,\, \lambda_2/\omega\gg 1 \,.
    \end{split}
\end{align}
The synchronization error of the network depends on the inverse of $(\tilde{\mathbb{L}} + {\bf K})$\,, which typically indicates that the effect of the attack spread across the network. Note that, when the edge weights to the attacked node are homogeneous, the first term in Eq.~(\ref{lf}) is simply the degree of node $k$ times the periodic function.
\begin{figure}
    \centering
    \includegraphics[scale=0.4]{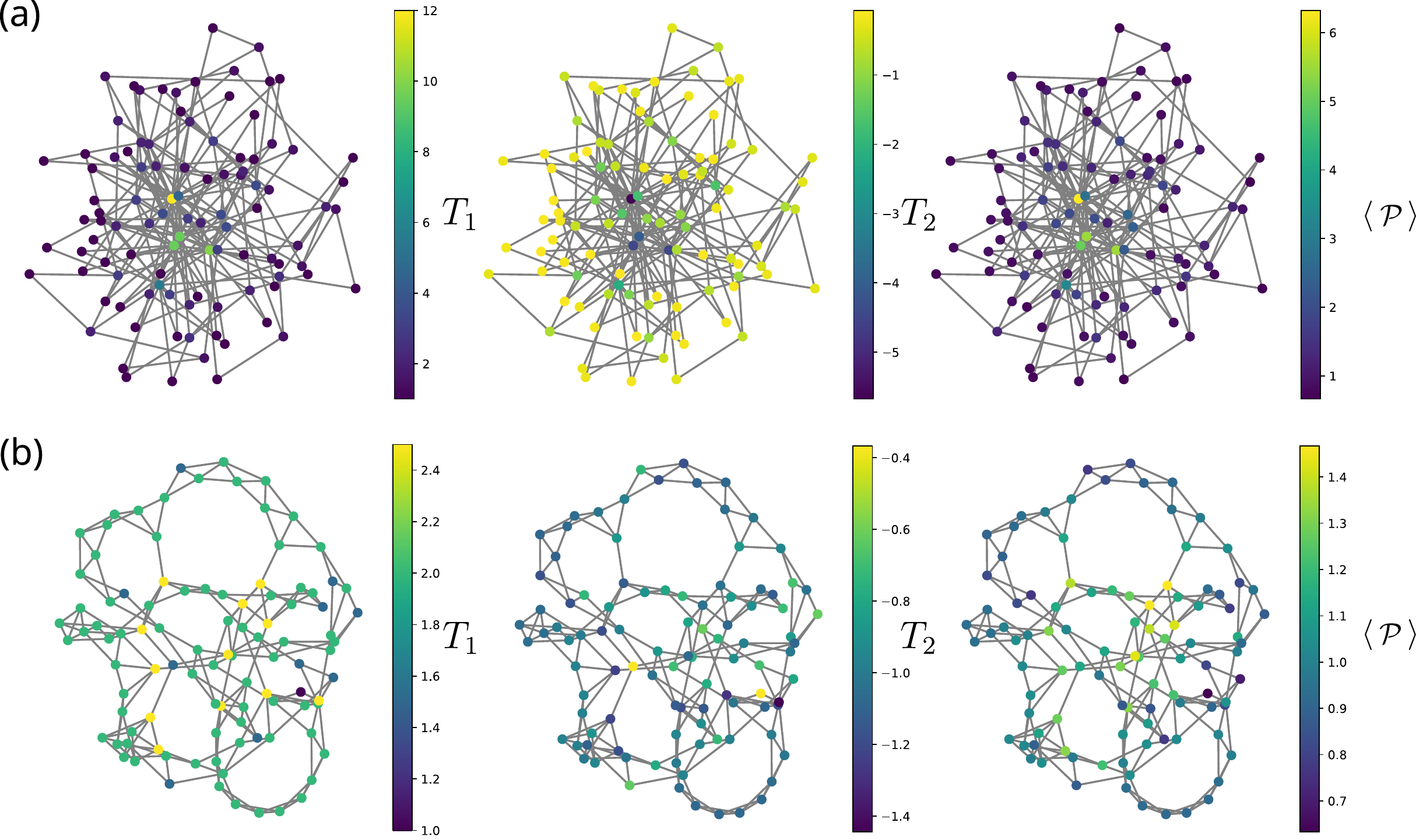}
    \caption{Color maps of $T_1$ (left panels), $T_2$ (middle panels) and the synchronization error (right panels) for (a) a Barab\'asi-Albert with $m=2$ and (b) a Watts-Strogatz network~\cite{New18book} with rewiring probability $p=0.05$\,, 4 initial nearest-neighbors and $N=100$\,. The weights on the edges are homogeneous. One observes that, as expected from Eqs.~(\ref{numeq}), (\ref{numeq2}), $T_1$ is given by local properties around the attacked node, while $T_2$ seems to depend not only on the local structure. The synchronization error is given by the sum of $T_1$ and $T_2$\,.}
    \label{fig4}
\end{figure}

\section{Numerical simulations}\label{Num}
For simplicity, we consider in this section only systems where all oscillators have the same natural frequency i.e. $\omega_i=0$ $\forall i$\,, which means that one of the synchronous states satisfies $\theta_i^{(0)}=C$ $\forall i$\,, where $C\in\mathbb{R}$ is some constant. 

In the previous section, we derived a closed form expression for the synchronization error Eq.~(\ref{eq11}) when the input signal is a white noise. The first term in this expression is simply proportional to the sum of the squared weights of the attacked node $k$\,. Here we illustrate numerically this results and show how both terms in Eq.~(\ref{eq11}) are important to assess the potential of an attack. Let us denote the two terms as,
\begin{eqnarray}\label{numeq}
    T_1 &=& \frac{\tau_0}{2}\sum_{j\in\mathcal{N}(k)} \tilde{b}^2_{jk} \,,\\ 
    T_2 &=& -\tau_0\sum_{\alpha,\beta}\sum_{i,j,l\in\mathcal{N}(k)}\tilde{b}_{ik}\tilde{b}_{jk}\tilde{b}_{lk}\frac{u_{\alpha,i}u_{\beta,j}u_{\alpha,l}u_{\beta,l}}{\lambda_\alpha+\lambda_\beta}\,.\label{numeq2}
\end{eqnarray}
Note that $T_1$ is computationally easier to obtain compared to $T_2$\,. Figure~\ref{fig4} shows $T_1$ (left panels), $T_2$ (center panels) and $\langle \mathcal{P} \rangle$ (right panels) when each node is attacked, for (a) a Barab\'asi-Albert and (b) a Watts-Strogatz network~\cite{New18book}. One observes that in Fig.~\ref{fig4}(a), the synchronization error is mostly given by the sum of the squared weights of the attacked node. This is due to the degree distribution in the network which is heterogeneous and varies over a relatively large interval. For such networks, one can identify the most vulnerable node as the one with the largest $T_1$\,. The situation is different in Fig.~\ref{fig4}(b), where the network has a more homogeneous degree distribution. Indeed, one observes that $T_1$ and $T_2$ are distinct enough such that one needs to consider the full expression $T_1+T_2$ to correctly identify the most vulnerable node.
The detailed answer given by Eq.~(\ref{eq11}) is therefore necessary to reliably identify the most vulnerable oscillators. This result is particularly interesting as it might appear as counter intuitive. Indeed, one could think that attacking the node with the highest degree is obviously the most efficient strategy. However, while this might be true in specific cases where the degree is widely distributed so that a small fraction of nodes have degrees much larger than all the others (i.e. $T_1\gg T_2$ for such nodes with very high degree), it is not true in general. For some networks, the degree does not vary much among the different nodes e.g. the network in Fig.~\ref{fig4}(b). Moreover, using the configuration model, one could build many different instances of networks where all the nodes have the same degree yet their connectivities are different~\cite{New18book}. One should therefore use Eq.~(\ref{eq11}) to correctly identify vulnerabilities to a Byzantine attack injecting a random signal.
\begin{figure}
    \centering
    \includegraphics[scale=0.42]{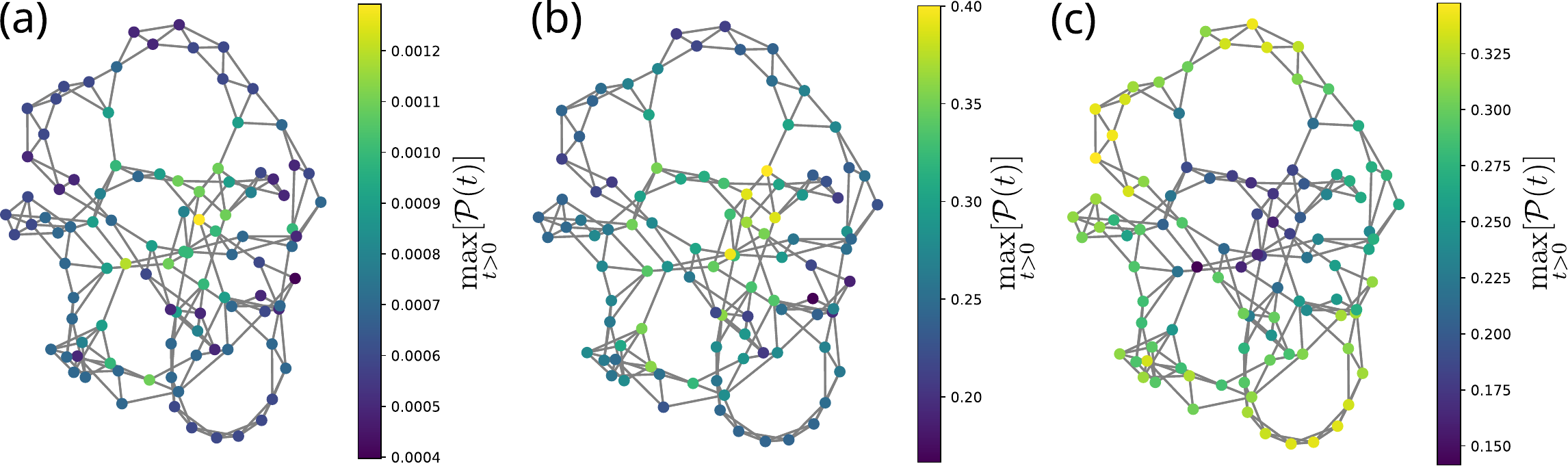}
    \caption{Color maps of the maximum of the synchronization error for the periodic input signal given in Eq.~(\ref{forc}) and different values of $\omega$\,. (a) High-frequency regime $\lambda_N/\omega\cong 0.072$\,; (b) intermediate regime $\lambda_N/\omega\cong 0.36$\,; (c) low-frequency regime $\lambda_2/\omega\cong 91.68$\,. As expected from Eqs.~(\ref{hf}), (\ref{lf}), the synchronization error in  (a) the high-frequency regime is given by local properties around the attacked node, while in both (b) the intermediate and (c) the low-frequency regimes, it seems to depend not only on the local structure.}
    \label{fig5}
\end{figure}
To further illustrate how to assess the vulnerability of the network to Byzantine attacks, we numerically investigate the periodic input presented in the previous section.
Figure~\ref{fig5} shows $\max_{t>0}[\mathcal{P}(t)]$ for the same Watts-Strogatz network as in Fig.~\ref{fig4}(b) and for different regimes of $\omega$\,. In Fig.~\ref{fig5}(a), one has the high-frequency regime [see Eq.~(\ref{hf})] where the synchronization error depends on the local structure of the network around the attacked node. The low-frequency case is given in Fig.~\ref{fig5}(c) [see Eq.~(\ref{lf})] that depends on the inverse of the matrix $[\tilde{\mathbb{L}} + {\bf K}]$\,, which is a more global quantity. One remarks that the most vulnerable nodes in panel (a) and (c) are different. The intermediate situation where $\omega$ is within the spectrum of $[\tilde{\mathbb{L}} + {\bf K}]$ is shown in Fig.~\ref{fig5}(b). Here the most vulnerable node is again different compared to the other regimes. Note that choosing $\omega$ within the spectrum of $[\tilde{\mathbb{L}} + {\bf K}]$ gives a larger synchronization error than the two limiting cases. Overall, comparing the right panel of Fig.~\ref{fig4}(b) and Fig.~\ref{fig5} indicates that the vulnerability to Byzantine attacks can vary significantly, depending on the type of perturbation and the intrinsic time-scales of the system.

\section{Conclusions}\label{conc}
We defined a framework for the analysis of Byzantine attacks on phase-coupled oscillators. We derived a closed form expression for the linear response of the system around its initial fixed point, to a general input signal from the attacker. The latter depends on the eigenvectors and eigenvalues of a matrix that is the sum of the Laplacian matrix of the network where the attacked node has been removed, and a diagonal matrix encoding the influence on the neighboring nodes. Interestingly, such systems has been considered in the field of opinion dynamics. Based on this framework, we then assessed the global impact of such attacks on the synchronous state using the synchronization error. One should note that other performance metrics can be computed within this framework. As a simple application of our theory, we first considered attacks where the degree of freedom of a single oscillator is replaced by a random input signal uncorrelated in time. In this case, the synchronization error can be expressed in terms of the sum of the squared weights of the attacked node plus a term that depends on the localization of its neighbors on the eigenmodes of the matrix $(\tilde{\mathbb{L}} + {\bf{K}})$\,. Second, we treated the case of a periodic input signal. In the high-frequency regime, the synchronization essentially depends on the local structure of the network close to the attacked node, while in the low-frequency regime, the response depends more globally on the network through the matrix $(\tilde{\mathbb{L}} + {\bf{K}})$\,. We observed that the vulnerability of the system crucially depends on the properties of the input signal and the coupling network.
The framework we presented here gives an efficient method to identify the most vulnerable and most robust oscillators in the system subjected to this type of attack, and therefore allows to take preventive measures in protecting and monitoring those that are vulnerable.

Based on previous results on the robustness of networked oscillators to noise inputs in the natural frequencies, it is expected that the synchronization error will depend on network properties that are more global than $T_1$ if the time-correlation of the attacker's signal is within or slower than the time-scales of the system~\cite{Tyl18a}.
Direct extensions of these results should consider more complicated input signals such as correlated noise, or copies of some other degrees of freedom in the system, and also include budget constraints for the attacker. One should notice that our theory can easily account for multiple input signals, corresponding to attacks where multiple oscillators are controlled at the same time.

\section*{Acknowledgements}
We thank A. Lokhov and M. Vuffray for useful discussions. This work has been supported by the Laboratory Directed Research and Development program of Los Alamos National Laboratory under project numbers 20220797PRD2 and 20220774ER and by U.S. DOE/OE as part of
the DOE Advanced Sensor and Data Analytics Program.
\providecommand{\newblock}{}

\end{document}